\documentclass[10pt, conference, letterpaper]{ IEEEtran}
\IEEEoverridecommandlockouts
\usepackage{cite}
\usepackage{amsmath,amssymb,amsfonts}
\usepackage{algorithmic}
\usepackage{graphicx}
\usepackage{textcomp}
\usepackage{algorithm}
\usepackage{booktabs}
\usepackage{xcolor}
\usepackage{multirow}
\usepackage{array}
\usepackage{colortbl}
\usepackage[T1]{fontenc}
\usepackage{times}           
\usepackage{mathptmx}        
\usepackage{textcomp}
\usepackage[hidelinks]{hyperref}

\def\BibTeX{{\rm B\kern-.05em{\sc i\kern-.025em b}\kern-.08em
    T\kern-.1667em\lower.7ex\hbox{E}\kern-.125emX}}
\begin{document}

\title{PRIME: Plasticity-Robust Incremental Model for Encrypted Traffic Classification in Dynamic Network Environments\\
{\footnotesize}
}

\author{\IEEEauthorblockN{1\textsuperscript{st} Tian Qin}
\IEEEauthorblockA{\textit{Southeast University} \\
Nanjing, China \\
230208959@seu.edu.cn}
\and
\IEEEauthorblockN{2\textsuperscript{nd} Guang Cheng}
\IEEEauthorblockA{\textit{Southeast University} \\
Nanjing, China \\
chengguang@seu.edu.cn}
\and
\IEEEauthorblockN{3\textsuperscript{rd} Zihan Chen}
\IEEEauthorblockA{\textit{Southeast University} \\
Nanjing, China \\
zhchen@njnet.edu.cn}
\and
\IEEEauthorblockN{4\textsuperscript{th} Yuyang Zhou}
\IEEEauthorblockA{\textit{Southeast University} \\
Nanjing, China \\
yyzhou@seu.edu.cn}
}

\maketitle

\begin{abstract}
With the continuous development of network environments and technologies, ensuring cyber security and governance is increasingly challenging. Network traffic classification(ETC) can analyzes attributes such as application categories and malicious intent, supporting network management services like QoS optimization, intrusion detection, and targeted billing. As the prevalence of traffic encryption increases, deep learning models are relied upon for content-agnostic analysis of packet sequences. However, the emergence of new services and attack variants often leads to incremental tasks for ETC models. To ensure model effectiveness, incremental learning techniques are essential; however, recent studies indicate that neural networks experience declining plasticity as tasks increase. We identified plasticity issues in existing incremental learning methods across diverse traffic samples and proposed the PRIME framework. By observing the effective rank of model parameters and the proportion of inactive neurons, the PRIME architecture can appropriately increase the parameter scale when the model's plasticity deteriorates. Experiments show that in multiple encrypted traffic datasets and different category increment scenarios, the PRIME architecture performs significantly better than other incremental learning algorithms with minimal increase in parameter scale.
\end{abstract}


\section{Introduction}
The continuous development of Internet technology has given rise to emerging services such as short videos and online payments, deeply integrating into and changing people's lives. In recent years, with the maturity of cloud computing technology and large language model, new directions have emerged for Internet architecture, but this has also posed new challenges for network management~\cite{LLMnetworking}. Network traffic classification can analyze various attributes of the traffic generated by communication between devices, including application, service type, VPN usage~\cite{vpn}, and whether there are malicious attacks~\cite{malware}, through port mapping or traffic monitoring devices. This analysis provides intelligence support for optimizing QoS/QoE~\cite{khanna2006application}, intrusion detection, and service-specific billing by ISP~\cite{ISP}, thereby enhancing the timeliness and effectiveness of network management services.\par
However, with the growing enhancement of internet security protection and privacy awareness, the TLS protocol has emerged, leading to the development of a series of encryption protocols (such as HTTPS, QUIC, etc.) that cater to various services. According to the Google report~\cite{googletransparencyreport} in April 2023, the proportion of encrypted webpages within the Chrome browser reached 98$\%$. While encrypted network traffic provides reliable privacy protection, it also presents certain challenges for network management. Traditional traffic classification techniques based on plain-text analysis are gradually becoming ineffective. Meanwhile, the emergence of proprietary protocols and the ever-changing communication patterns resulting from application layer encryption significantly reduce the applicability of classification techniques based on simple rules or machine learning methods, such as those relying on ports and statistical features~\cite{2010Automatic}.\par
Network traffic, as a standardized sequence of interaction data, retains characteristics such as data types and sizes that influence the segmentation of transmitted data frames, even if the content being transmitted is obscured by encryption. Additionally, the directionality of data transmission and the frequency of interactions within the bi-flow can reflect certain patterns that are useful for traffic classification. Therefore, by extracting the packet length sequence from the transport layer data and combining it with the directionality and temporal frequency of packet transmission, we can obtain valuable information. Additionally, incorporating unencrypted protocol header provides more sufficient information gain for traffic classification. Building upon these rich feature sets, recent advances in deep learning architectures like transformers~\cite{lin2022bert} have shown remarkable effectiveness in processing sequential network traffic data. Moreover, by integrating data augmentation techniques like contrastive learning, these models can further accommodate challenging real-world conditions such as packet loss and retransmissions~\cite{Rosetta}.\par
These sophisticated models, however, face a fundamental challenge: network traffic data exists in a constantly changing environment. Emerging services, proprietary protocols in new scenarios, continuously iterating application versions, and the emergence of new types of attacks can render existing classification models ineffective, leading to what is known as "concept drift"~\cite{drift}. For example, new Internet of Things (IoT) protocols can introduce new reflection pathways that result in novel DDoS attacks, which may bypass existing traffic classification models used for intrusion detection. To handle concept drift, defenders must typically reconstruct training datasets and retrain their models entirely. As model complexity increases, this retraining process becomes increasingly time-consuming, creating dangerous delays in responding to emerging attack patterns.\par
To ensure the real-time effectiveness of network traffic classification models, researchers typically adopt incremental learning approach to enable rapid adaptation to new samples. Incremental learning reuses most of the parameters from the existing model to maintain recognition performance for the original category samples. On this basis, model fine-tuning is performed on several network layers at the end to achieve recognition of new types of samples. However, this approach often faces the issue of "catastrophic forgetting," where the fine-tuned model may significantly degrade in performance on the original tasks, especially when there are many categories in the classification task. To address such issues, several common strategies have been proposed: 1. Replaying a representative subset of the original category samples to prevent the model's update direction from drifting; 2. Introducing a constraint loss function that penalizes excessive changes in model weights; 3. Separating the parameters used for classifying new category tasks to ensure they do not affect the performance of the original tasks.\par
The first two types of methods can incur additional overhead in model training, and the effectiveness of the improvements is related to factors such as sample distribution, differences between samples, and the total number of sample categories, which makes them less deterministic. Although the approach of separating model parameters can reliably accommodate the classification performance of both original and new categories, it increases the scale of the model, resulting in additional computational and storage overhead during model deployment. Moreover, traffic classification tasks are often high-density, high-concurrency processes, and at critical nodes in the backbone network, the throughput of traffic data can reach 100 Gbps ~\cite{Gbps}. Additionally, ensuring the normal operation of user services is a fundamental requirement across all scenarios. However, in environments like the Internet of Things (IoT), where devices often have limited computational power and storage resources, it becomes especially critical for network traffic classification tasks to minimize resource consumption. Therefore, the parameter scale of models used for network traffic classification should not be blindly expanded in pursuit of classification accuracy. Thus, we aim to integrate the above ideas and separate model parameters at appropriate times to achieve efficient model evolution.\par
The challenges of adaptability brought about by continuous learning naturally attract the attention of researchers. In 2024, Shibhansh Dohare et al. proposed a study on the plasticity properties of neural network models, providing an effective theoretical basis for explaining the phenomenon of "catastrophic forgetting" that occurs in incremental learning scenarios~\cite{forgetting}. By observing the patterns of change in neuron parameter weights during incremental learning tasks, it was found that as new classification tasks are introduced, a large number of neurons become "stagnant," meaning their weights change very little. The decline in neuron plasticity partially explains the reasons for catastrophic forgetting; when the total number of tasks is too large, it becomes difficult for the model to accommodate different classification tasks. The introduction of neural network model plasticity provides an important metric for assessing whether the model's scale needs to be expanded, guiding the orderly evolution of network traffic classification models.\par
In response to the aforementioned issues, this article proposes the PRIME architecture. For the newly added traffic classification tasks, the LwF(Learning without Forgetting) algorithm is first employed for replay-free incremental learning. At the same time, the system detects the model's plasticity using two indicators: the proportion of inactive neurons (could be measured by entropy efficiency) and the effective rank of parameters. Once a decline in model plasticity is detected, the system implements a model expansion strategy to enhance the representational capacity of the model, ultimately training a new model with optimal parameter size and incremental effectiveness.\par
The organization of the remaining content in this paper is as follows: Section II discusses research related to network traffic classification and incremental learning, while Section III introduces the concept of neural network plasticity along with theoretical derivations of other related components. Section IV presents the overall design of the PRIME architecture and its associated operational mechanisms. In Section V, incremental learning scenarios are constructed using mainstream datasets, and the model evolution effects of the proposed architecture are compared with those of other algorithms. The final section summarizes the content of this paper and offers prospects for future work.

\section{Preliminaries}

\subsection{Feature Extraction and Deep Learning Methods for Encrypted Traffic}
Network traffic, as a standardized form of data, contains information at the transport and application layers that may indicate specific application types and behavioral patterns, including potential malicious activities. However, plaintext fields such as address information, port numbers, and SNI fields, which are susceptible to tampering or transformation, may lead to unnecessary associations with categorical attributes, resulting in model overfitting. Therefore, specific feature extraction methods are needed to capture attributes that are highly relevant to behavior. \par
We refer to a general feature processing approach for encrypted traffic~\cite{distiller}. The feature attributes of this traffic unit can be roughly divided into two categories: one derived from the transport-layer payload. The other from residual information such as protocols and transmission modes. The feature attributes of this traffic unit can be roughly divided into two categories: one derived from the transport-layer payload. The other from residual information such as protocols and transmission modes.\par
For the transport-layer payload, although its plaintext information is obscured by the TLS encryption protocol, the way a single flow segments its transmitted data and its interaction patterns are sufficient to establish a correlation with its class affiliation. Therefore, for a single flow $T$, we extract the first $N_b$ bytes of the transport-layer payload of the Traffic Classification object (instances longer than $N_b$ bytes are truncated; shorter instances are zero-padded to match $N_b$). The input is represented in binary format, organized in a byte-wise manner, and normalized within the range of $(0, 1)$. Let $X_{pay}$ denote the final extracted feature of the payload, and $b_i$ represent the byte representation of the payload content. The relationship can be expressed as follows:
$$X_{pay} = \left[ b_1, b_2, \ldots, b_N \right]/255, \quad b_i \in [0, 255] $$ \par

The interaction of network traffic exhibits a certain Markov property, whereby the characteristics of each data packet—such as its size, direction, and arrival time—are influenced by the transmission behavior of the preceding packet. This intrinsic property is determined by specific behavioral patterns and categories of applications that influence the traffic flow. Therefore, in this study, we extract the following features from the first $N_p$ packets to facilitate classification: (i) the number of bytes in the transport-layer payload, (ii) the TCP window size (set to zero for UDP packets), (iii) the inter-arrival time, and (iv) the packet direction $\in{0,1}$.\par

After extracting the features, we employed the deep learning supervised paradigm to train the traffic classification model. Assuming basic traffic unit $T(X, Y)$ for classification, where $X = [X_{pay}, X_{hdr}]$ represents the feature vector extracted from this sample, which is the combination of the two types of features mentioned above, and $Y$ is the class label of the sample. For supervised training, the objective is to learn a mapping function $F$ to predict $\hat{Y}$ from $X$. Then for the model parameter $\theta$, we have the following equation  $\hat{Y} = F(X;\theta)$.\par
In the regular training step of this work, the model is trained by minimizing the cross-entropy loss:
$$ L(\hat{Y}, Y) = -\sum_{i=1}^{C} Y_{i} \log(\hat{Y}_{i}) $$
Then, the parameters $\theta$ are updated using gradient descent: $\theta \leftarrow \theta - \alpha \nabla_{\theta} L$. Through this process, we enhance the model's capability to classify traffic flows accurately.

\subsection{Incremental Learning and Model Plasticity Analysis}
In a constantly changing network environment, traffic classification models often face samples such as application iterations and new types of attacks. This necessitates that the model, while relying as little as possible on old task data, uses incremental learning methods based on new task data to train and enhance the model, achieving a globally optimal state that accommodates classification tasks for both new and old sample spaces.\par
Assuming the current model is facing a new traffic classification task $T_{n+1}$, after having already completed several tasks $T_1 \sim T_n $, the current incremental learning goal is to find the model parameters $\theta^*$ that satisfy:
$$\sum_{j=0}^{n+1} \mathbb{E}_{(X^{T_j}, Y^{T_j})} [L(F_j(X^{T_j}, \theta^*))]  $$
In the scenario of iterating on traffic classification models, replaying old data implies the need for additional storage and computing resources. Therefore, it is necessary to partition the model, training additional parameters with new data while employing certain regularization methods to prevent excessive shifts in the model’s predictions for old tasks. Thus, the model can be roughly divided into four parts: $\theta_z$ for the frozen parameter layer, $\theta_s$ for the shared parameter layer, and $\theta_{T_{n+1}}$ and $\theta_o$ for the classification layers related to the new and old tasks, respectively. Here, we introduce the LwF algorithm. First, we calibrate the initial model $\theta_1=[\theta_z,\theta_s,\theta_o]$ to obtain the predicted output $Y^{T_n+1}_o$ for the new task. Then, when updating $\theta_s$ and $\theta_{T_{n+1}}$ through supervised training, we add an additional penalty based on the deviation of the output from $\theta_o$ relative to $Y^{T_n+1}_o$, corresponds to an improved cross-entropy function: 
$$ L^*(\hat{Y}^{T_n+1}_o, Y^{T_n+1}_o) = -\sum_{i=1}^{C} \frac{(y_o^{(i)})^{1/T}}{\sum_{j}(y_o^{(j)})^{1/T}} \log \frac{(\hat{y}_o^{(i)})^{1/T}}{\sum_{j}(\hat{y}_o^{(j)})^{1/T}} $$
The overall update objective of the model $\theta^* = [\theta_z, \theta_s^*, \theta_o, \theta_{T_{n+1}}^*]$ can be formulated as:
$$\mathop{\operatorname{argmin}}_{\hat{\theta}_s,\hat{\theta}_{T_{n+1}}} (\lambda_0 L^*(\hat{Y}^{T_n+1}_o, Y^{T_n+1}_o) + L(\hat{Y}^{T_n+1}, Y^{T_n+1}) + R(\hat{\theta})) $$\par
However, although deep learning models are mostly over-parameterized, the capacity for model recognition is ultimately limited, and the LwF method is highly constrained by the correlations between tasks. Therefore, this work will perform a plasticity assessment of the model after the LwF update to determine whether there is a need to expand the model's parameter scale based on its current state. According to relevant research ~\cite{dohare2024loss}, a decline in plasticity can lead to poor performance of the model on incremental tasks, and the literature has also proposed two approaches to measure the plasticity of the model.\par
We find that entropy-based metrics effectively capture the proportion of inactive neurons in our progressive expansion scenario, as entropy values reflect the diversity and utilization of neuronal representations. Therefore, our first approach evaluates network capacity through an information density measure derived from the entropy of neural activations.\par
For neurons in shared parameters $\theta_s$, we first compute the Shannon entropy $H$ of their activation distribution, then derive an information density measure as: $E_s = \frac{H}{n^\alpha}$. In this equation, $H = -\sum_{i=1}^{k} p_i \log_2(p_i)$ represents the Shannon entropy of the activation distribution, $n$ is the current neuron count, and $\alpha$ is a scaling parameter that controls the sensitivity to network scale versus representation granularity. For computational efficiency, we utilize the distribution of $L_\infty$ norms of neural activations as a proxy for the activation distribution.\par 
The measure $E_s$ quantifies the information density of neural representations by capturing how efficiently the network utilizes its neuronal capacity to encode diverse activation patterns. A lower value indicates underutilized capacity with significant potential for expansion and performance improvement, while a higher value suggests the network is approaching its representational limits with limited room for further enhancement.\par
The second approach is to measure the effective rank of the model's parameter matrix. Research indicates that if the effective rank of a particular layer in a neural network is low, it means that the output of that layer can be derived from a small number of neurons. This pattern also suggests a decrease in the model's plasticity~\cite{7098875}.\par
Due to the computational complexity involved in singular value decomposition, this paper randomly selects a specific layer $l^*$ of the parameters $\theta_s$ for effective rank calculation each time. Let the singular values of the parameter matrix of layer $l^*$ be $\sigma_1 \geq \sigma_2 \geq \cdots \geq \sigma_r$. We filter out singular values smaller than threshold $\epsilon = 1 \times 10^{-5}$ to ensure numerical stability.
The formula for calculating the effective rank is:
$$R_e = \exp\left(-\sum_{i: \sigma_i > \epsilon} \frac{\sigma_i}{\|\sigma\|_1} \log \frac{\sigma_i}{\|\sigma\|_1}\right)$$
where $\|\sigma\|_1 = \sum_{j: \sigma_j > \epsilon} \sigma_j$.

\section{System Overview}
In this paper, we divided the operational cycle of the encrypted traffic classification model into three phases. The first phase involves training a model in a static network topology, making it difficult to recognize emerging traffic samples in real world, which leads to obsolescence. Subsequently, incremental learning is performed using new samples to achieve model iteration. Finally, the model reaches an ideal state that is suitable for the current environment, demonstrating optimal usability alongside a well-balanced parameter scale. These three states are illustrated in Figure 1. The PRIME architecture proposed in this article is designed to ensure optimal efficiency throughout the entire incremental learning process in the model iteration, achieved through a structured four-step approach.\par
Remarkably, none of the steps in the PRIME method require replaying old task data, saving storage costs. Moreover, we deconstruct the model for encrypted traffic classification into Fine-tune layers, Hidden layers, and the final task layer. In our approach, only the parameters in the Hidden layers and the final task layer are updated, while most parameters remain frozen, resulting in significant savings in computational resources. In the first step (A), we will employ the LwF algorithm to conduct incremental learning on the model's final task layer, followed by a plasticity analysis of the model slices adjacent to the task layer in step (B). If the plasticity is up to standard, the update process concludes, resulting in a simpler incremental update. Otherwise, we will move to step (C) to formulate an appropriate model expansion strategy, approaching the Net2Net~\cite{chen2015net2net} method to reasonably expand the parameter size of the model. Finally, in step (D), by combining the LwF and Net2Net algorithms for training in new task scenarios, a plasticity-healthy incremental model is obtained. Detailed explanations of each step can be found in the respective subsections of this section.

\begin{figure*}[h]
	\centering
	\includegraphics[width=0.99\textwidth]{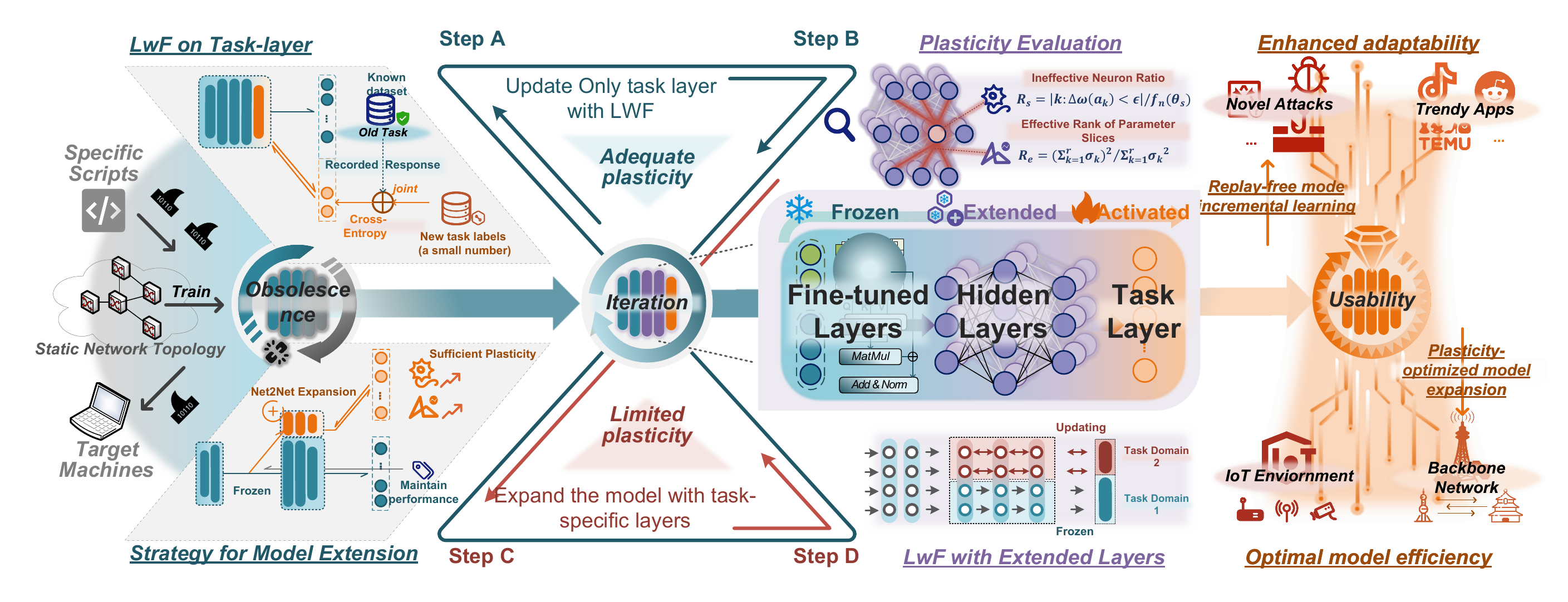} 
	\caption{Overview of the PRIME System Structure \\
		\small This figure illustrates the process of transitioning the encrypted traffic classification model from obsolescence to usability using the PRIME approach, with most parameters frozen. There are four core steps involved: 1) LwF on Task Layer, 2) Plasticity Evaluation, 3) Strategy for Model Extension, and 4) LwF with Extended Layers. When model plasticity is sufficient, only the first half of the cycle will be executed; otherwise, the complete cycle will be performed to expand the parameter scale.}
	\label{fig1}
\end{figure*}

\subsection{LwF on Task Layer}
For a small number of new classification tasks, since neural network models are often over-parameterized, a new task layer can be directly added to learn the new classification task. The parameters of the new task layer and the shared layers use the labels of the new sample, applying the cross-entropy function to compute the loss and updating parameters using gradient descent. Certainly, as mentioned in the Preliminaries, the LwF algorithm should be used to avoid the catastrophic forgetting problem. Therefore, during the initial training of the new data samples, our system records the output logits from the original task layer as calibration values. During the training process, the previously mentioned constraint loss function $L^*$ is trained in conjunction with the cross-entropy loss. The specific process of this step is shown in Algorithm 1.

\begin{algorithm}
	\caption{LwF Training on Task Layer}
	\label{alg:new_task_training}
	\begin{algorithmic}[1]
		\REQUIRE new\_tasks $T_{n+1}$ with $T^{T_{n+1}}(X^{T_{n+1}},Y^{T_{n+1}})$\\
		origin model: $\theta_1=[\theta_z,\theta_s,\theta_o]$
		\STATE \textbf{Notation:} learning rate: $\alpha$, max epochs: $t_{max}$, class num of $T_{n+1}$: $ C_{n+1}$, joint loss function: $\tilde{L}$, the $t_{th}$ round parameters: $\theta^{(t)}$ \\ The other symbols are the same as in the main text. 
		\STATE Initialize model: $\theta_1=[\theta_z,\theta_s,\theta_o] \gets \text{init\_model()}$
		\STATE Create new task layer: $\theta_{T_{n+1}} $
		\STATE Record calibration values: $\hat{Y}^{T_n+1}_o \gets F_n(X^{T_{n+1}}, \theta_1) $
		\STATE Add $\theta_{T_{n+1}}$ to $\theta_1$ : $\theta^{(0)} = [\theta_z,\theta_s,\theta_o, \theta_{T_{n+1}}]$
		\FOR{epoch: $t$ = 1 to $t_{max}$}
		\STATE Predicted value: $\hat{Y}^{T_n+1} \gets F_{n+1}(X^{T_{n+1}}, \theta^{(t)})$
		\STATE Cross-entropy:  \vspace{-1.2em}
		\begin{align*} L(\hat{Y}^{T_{n+1}}, Y^{T_{n+1}}) = -\sum_{i=1}^{C_{n+1}} Y^{T_{n+1}}_{i} \log(\hat{Y}^{T_{n+1}}_{i})  \end{align*} \vspace{-1.2em}
		\STATE Joint loss: \vspace{-0.2em} \small
		\begin{align*}  \tilde{L} \gets \lambda_0 L^*(\hat{Y}^{T_n+1}_o, Y^{T_n+1}_o) + L(\hat{Y}^{T_n+1}, Y^{T_n+1}) + R(\hat{\theta}^{(t)}) \end{align*} \vspace{-1.2em}
		\STATE Gradient descent: \vspace{-0.2em}
		\begin{align*} \theta^{(t)}_s \leftarrow \theta^{(t-1)}_s - \alpha \nabla \tilde{L} ; \quad \theta^{(t)}_{T_{n+1}} \leftarrow \theta^{(t-1)}_{T_{n+1}} - \alpha \nabla \tilde{L} \end{align*}  \vspace{-1.2em}
		\STATE Overall parameters: $\theta^{(t)} = [\theta_z,\theta^{(t)}_s,\theta_o, \theta^{(t)}_{T_{n+1}}]$
		\ENDFOR
		\STATE Training completed: $\theta^* = [\theta_z,\theta^{(t_{max})}_s,\theta_o, \theta^{(t_{max})}_{T_{n+1}}]$
	\end{algorithmic}
\end{algorithm}

\subsection{Plasticity Evaluation}
Although over-parameterized models allow for some degree of continued compatibility with incremental tasks based on the existing parameter scale, the original model will gradually reach its limit in representational complexity as the number of tasks increases. This will lead to a decrease in plasticity. In this case, there will be an irreconcilable contradiction between the model performance on new tasks and that on existing tasks. Therefore, during the actual operation of the PRIME system, when the simple LwF method results in the classification performance of new tasks being below expectations (specifically manifested as the cross-entropy loss of the new tasks oscillating significantly above a defined value), two methods will be employed to assess plasticity: measuring the distributional diversity of neural activations and calculating the effective rank of parameters. After a comprehensive evaluation, it will be determined whether the current model has limited plasticity. Once plasticity falls below a threshold, the PRIME system will immediately halt the ongoing incremental learning and proceed with the formulation of model expansion strategies. Hence, this process occurs interspersed within Step A; for brevity, it is not shown in Algorithm 1. The effective rank of the parameter matrix in a certain layer of the neural network being closer to the dimension of that layer somewhat indicates that the feature representation capability of that layer has reached its limit. Similarly, an entropy efficiency $E_s$ approaching 0 also indicates that the model's training has stagnated, as it reflects a highly uniform activation distribution where most neurons contribute minimally to the representational diversity. Therefore, combining the effective rank and the inactive neuron determination formula(the entropy efficiency) mentioned in section Preliminaries, we can derive the parameter $P$ that measures the plasticity of a layer with $n$ neurons as follows:
$$P =\omega_1 \frac{R_e}{n} + \omega_2 E_s = \omega_1 \cdot pr_1 + \omega_2 \cdot pr_2$$
where $\omega_1$ and $\omega_2$ are the weight coefficients that balance the importance of two indicators. After obtaining the ratios $pr_1$ and $pr_2$, the two can be combined to form a plasticity discriminative coefficient. Thus, we set a threshold $\kappa_s$. When the plasticity indicator exceeds this threshold, we believe that the model's expressive capability is lacking, and it is necessary to perform Step C.\par
The specific formulas for assessing plasticity and the selection of relevant thresholds need to be based on observations of the numerical changes during normal model training, and the detailed process is outlined in Section V.B.

\subsection{Strategy for Model Extension}
Limited representational capacity in incremental learning scenarios can lead to severe catastrophic forgetting. However, in the field of traffic classification, both high-throughput backbone network nodes and resource-constrained small devices in Internet of Things (IoT) or mobile communication contexts require compact models to ensure rapid processing rates and efficient utilization of limited computational resources. Therefore, upon detecting limited plasticity, it is essential to implement appropriate strategies to expand the model's scale. The PRIME architecture employs the Net2Net algorithm to increase the model's width. This algorithm facilitates the expansion of our model’s width by creating a new, larger network that replicates the original network’s parameters and add a slightly perturbation $\epsilon_0$. Assuming we expand a certain layer of the model to $r$ times its original size, the parameter matrix for this layer $\omega_{new} \in R^{m \times n}$ can be expressed as: $\omega_{new} = [\omega_{old}, \omega_{copy} + \epsilon_0]\in R^{m \times rn}$, where $\omega_{copy}$ contains duplicated columns from $\omega_{old}$. Due to the presence of perturbations, the newly added neurons are all in an active state, which increases the overall activation diversity and affects the entropy efficiency measurement. We use normalized entropy $E_s$ to measure parameter efficiency. This metric provides a principled way to evaluate information density while accounting for model complexity. The scaling behavior makes it suitable for comparing efficiency between models of similar scale and architecture.\par 
For a neural network layer with parameter matrix $\omega_i \in m \times n$ that has been expanded by a factor of $r$, the change in its effective rank proportion can be expressed as: $pr_2^* = \frac{R_e^*}{rn} = \frac{R_e + \Delta R}{rn}$.\par
The random disturbance we added follows independent and identically distributed Gaussian random variables. Considering the relevant inference by matrix perturbation theory ~\cite{li2006matrix}, the expected rank increment $\Delta R \approx min((r-1)n, n-R) $, where $R$ is the rank of the original parameter matrix. When we handle the traffic classification task, the number of fully connected layers in the task layer is generally greater than 32 dimensions, i.e., $n>32$. Therefore, the increment in rank is relatively small, so we can approximately consider that the effective rank proportion of the model is linearly related to the scale of model expansion. \par 
After clarifying that the two evaluation metrics are approximately linearly correlated with the expansion scale, the setting of the expansion scale $r$ is related to the plasticity threshold that triggers model expansion. It is necessary to set an appropriate expansion ratio while avoiding excessive model expansion or frequent triggering of expansion instructions. The selection of related parameters is based on this strategic standard, which is specifically elaborated in Section V.B.

\subsection{LwF with Extended Layers}
This step is similar to the LwF algorithm in Step A, but the parameters of the model before expansion will be frozen to fully protect the recognition performance of the old tasks. The newly added model parameters will be trained and updated on the new tasks according to the LwF algorithm. In summary, we can obtain a complete process for incremental learning under model expansion after triggering the plasticity threshold, as shown in Algorithm 2.

\begin{algorithm}
	\caption{LwF Training with Model Extension via Net2Net}
	\label{alg:new_task_training_with_net2net_extension}
	\begin{algorithmic}[2]
		\REQUIRE new\_tasks $T_{n+1}$ with $T^{T_{n+1}}(X^{T_{n+1}},Y^{T_{n+1}})$\\
		origin model: $\theta_1=[\theta_z,\theta_s,\theta_o]$
		\STATE \textbf{Notation:} Expansion scale: $r$, expanded parameters: $\theta_{E}$, parameter duplication: $ \theta_{copy}$\\ The other symbols are the same as in the main text and algorithm 1.
		\STATE Initialize model: $\theta_1=[\theta_z,\theta_s,\theta_o] \gets \text{init\_model()}$
		\STATE Model Extension: $\theta_{E} \gets \text{Net2Net}(\theta_s, \theta_{copy} \cdot (1-r) + \epsilon)$
		\STATE Create new task layer: $\theta_{T_{n+1}}$
		\STATE Freeze original parameters: $\theta_z, \theta_s, \theta_o$
		\STATE Record calibration values: $\hat{Y}^{T_n+1}_o \gets F_n(X^{T_{n+1}}, \theta_{E})$
		\STATE Combine parameters: $\theta^{(0)} = [\theta_z, \theta_{E}, \theta_o, \theta_{T_{n+1}}]$
		\FOR{epoch: $t$ = 1 to $t_{max}$}
		\STATE Predicted value: $\hat{Y}^{T_n+1} \gets F_{n+1}(X^{T_{n+1}}, \theta^{(t)})$
		\STATE Cross-entropy: \vspace{-1.2em}
		\begin{align*}
			L(\hat{Y}^{T_{n+1}}, Y^{T_{n+1}}) = -\sum_{i=1}^{C_{n+1}} Y^{T_{n+1}}_{i} \log(\hat{Y}^{T_{n+1}}_{i})  
		\end{align*} \vspace{-1.2em}
		\STATE Joint loss: \vspace{-0.2em} \small
		\begin{align*}
			\tilde{L} \gets \lambda_0 L^*(\hat{Y}^{T_n+1}_o, Y^{T_n+1}_o) + L(\hat{Y}^{T_n+1}, Y^{T_n+1}) + R(\hat{\theta}^{(t)})
		\end{align*} \vspace{-1.2em}
		\STATE Gradient descent: \vspace{-0.2em}
		\begin{align*}
			\theta^{(t)}_{E} \leftarrow \theta^{(t-1)}_{E} - \alpha \nabla \tilde{L} ; \quad \theta^{(t)}_{T_{n+1}} \leftarrow \theta^{(t-1)}_{T_{n+1}} - \alpha \nabla \tilde{L}
		\end{align*}  \vspace{-1.2em}
		\STATE Overall parameters: $\theta^{(t)} = [\theta_z, \theta^{(t)}_{E}, \theta_o, \theta^{(t)}_{T_{n+1}}]$
		\ENDFOR
		\STATE Training completed: $\theta^* = [\theta_z, \theta^{(t_{max})}_{E}, \theta_o, \theta^{(t_{max})}_{T_{n+1}}]$
	\end{algorithmic}
\end{algorithm}

\section{Experiment}
\subsection{Dataset Preparation and Experiment Settings}
In this paper, we utilize three datasets to comprehensively evaluate the detection performance of PRIME across different incremental learning scenarios. All three datasets focus on classifying the application attributes of network traffic, providing diverse challenges for incremental learning evaluation.\\
\textbf{IPTAS-Tbps}~\cite{chen2022a3c} consists of traffic from seven mainstream applications collected under CERNET, configured as a single increment scenario (4+3 classes) for evaluating basic task extension capabilities.\\
\textbf{ISCX-VPN2016}~\cite{VPN2016} contains traffic from applications under VPN and non-VPN scenarios, creating low task similarity conditions that represent challenging incremental learning scenarios.\\
\textbf{MIRAGE19}~\cite{aceto2019mirage} encompasses traffic from 40 Android applications, configured for continual increment evaluation with 10 base classes plus 2 classes per subsequent stage.\par
These datasets collectively enable comprehensive assessment of PRIME's performance across varying incremental learning complexities, from simple single-step extensions to complex multi-stage continual learning with different degrees of task relatedness.\par
In our experiments, we utilized deep learning models implemented with PyTorch 2.0.0 and CUDA 11.7. Data pre-processing and post-processing were primarily conducted using the NumPy and Scapy libraries. For graphical data representation, we employed Matplotlib and MATLAB. All experiments were performed on a PC with the following hardware specifications: a 13th Gen Intel® Core™ i7-13700KF processor running at 3.40 GHz, 32 GB of RAM, and an NVIDIA GeForce RTX 4080 GPU. The operating system used was Windows 11.\par

\subsection{Model plasticity detection and parameter selection}
To gain a more intuitive understanding of the relationship between model plasticity and detection performance, as well as to establish a reasonable plasticity threshold for the PRIME system, we first need to observe the entropy efficiency of inactive neurons and the distribution of parameter effective rank within the model parameters when they are at their limits. We have designed a basic model architecture based on the transformer to handle classification tasks on the three datasets. The transformer architecture~\cite{NIPS2017_3f5ee243} employs multihead attention composed of queries(Q), keys(K), and values(V), enabling it to extract sequential dependencies across varying spans, thus fulfilling the analytical requirements for both flow-level and packet-level modalities. The specific model structure parameters and hyper-parameters utilized for training each traffic classification task are presented in Table I.\par
\vspace{-2mm}
\begin{table}[ht]
	\centering
	\caption{\small Architecture and Hyper-parameters of Unified Models}
	\small 
	\begin{tabular}{@{}ll@{}}
		\toprule
		\textbf{Parameter} & \textbf{Value} \\ \midrule
		Transformer Encoder & 912dim, 2 heads \\ 
		Linear Layers & [912, 256, 64, $N_{\text{target}}$] \\ 
		Optimizer & Adam, 1e-3 \\ 
		LR Scheduler & ReduceLROnPlateau \\
		& (factor=0.5, patience=5) \\
		Activation & ReLU \\ 
		Dropout, Batch Size, Epochs & 0.2, 512, 30 \\ 
		Loss Function & Cross-Entropy \\ 
		\bottomrule
	\end{tabular}
	\label{tab:nn_hyperparameters}
\end{table}

We first aim to determine the relationship between the effective rank ratio of parameter matrices and model accuracy. Therefore, the experiment initially sets the Transformer layer as a shared pre-trained layer, followed by two fully connected operations serving as hidden layers to measure factors related to model plasticity, and finally connects a task layer of equivalent dimensions to output logits that directly provide classification information.\par
First, we deploy models with varying hidden layer sizes on the MIRAGE19 dataset, with the two fully connected layer dimensions ranging from 64 to 512. To facilitate observation of the impact of effective rank ratio on classification performance, we set the second hidden layer dimension uniformly to 64, then observe the classification performance when the first hidden layer dimensions are 64, 128, 258, 512, and 1024 respectively, and record the effective rank ratio of that layer when the model approaches convergence. The experimental results are shown in Figure 2. All experimental data in this paper are uniformly partitioned into 75$\%$ training set, 10$\%$ validation set, and 15$\%$ test set. All experimental results are obtained from 10 repeated experiments, with the dataset being randomly partitioned and shuffled for each experiment.\par
\begin{figure}[h]
	\centering
	\includegraphics[width=0.3\textwidth]{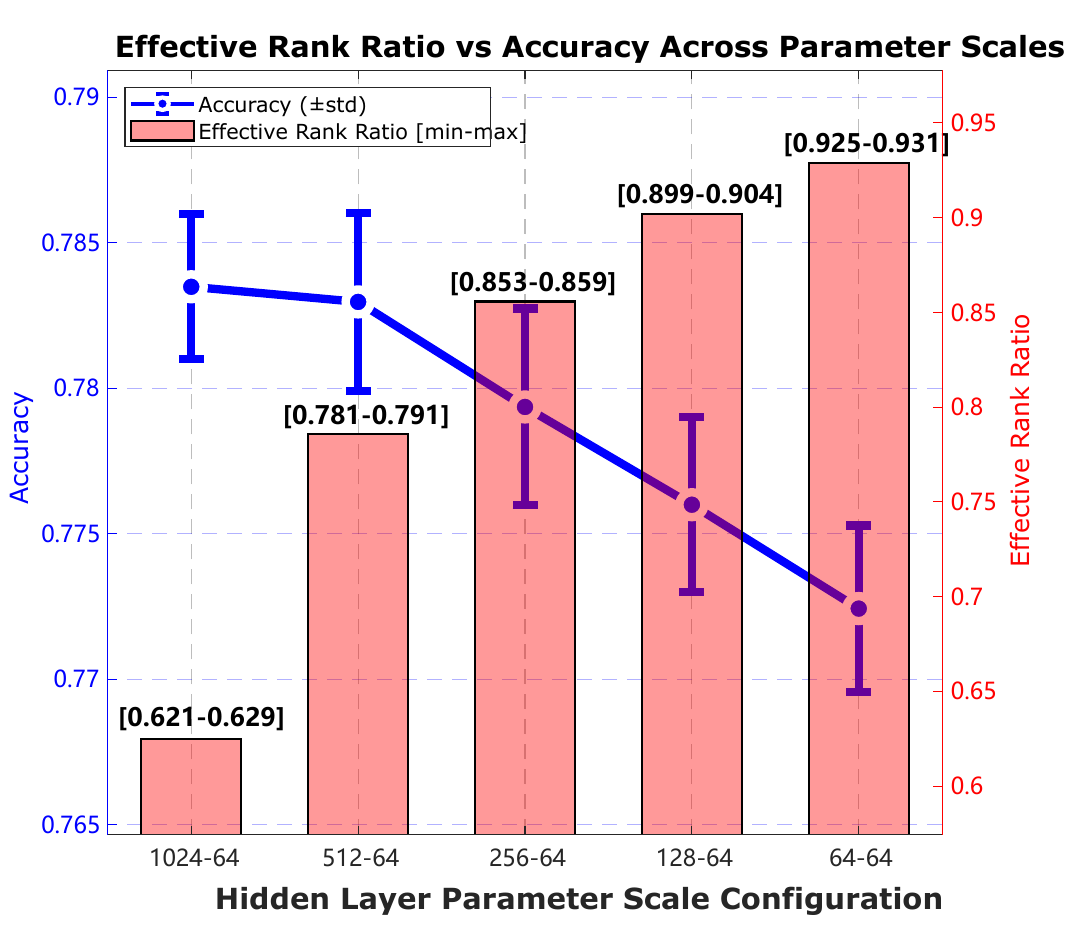} 
	\caption{Under conditions of limited model plasticity, recognition accuracy decreases as the effective rank ratio increases.}
	\label{fig2}
\end{figure}
The blue line in Figure 2 demonstrates the variation of model classification accuracy with model scale. It can be observed that when the number of neurons in the first layer is below 512, classification performance decreases as the number of neurons decreases, while beyond 512 dimensions, model classification performance shows no further improvement. Therefore, we can consider that when dimensions exceed 512, the model's representational capacity, i.e., plasticity, is sufficient, whereas when dimensions are at or below 256, plasticity is in a limited state. Through the effective rank ratio values in the bar chart, it is evident that the more insufficient the model plasticity, the closer this ratio approaches 1, which aligns with conclusions from related research and our aforementioned expectations. Combined with the accuracy results, we can approximately conclude that when this ratio (i.e., $pr_1$) exceeds 0.85, the model plasticity is in a limited state and requires triggering expansion strategies. Correspondingly, when this value is below 0.8, the model's representational capacity can be considered sufficient at that time point.\par
A single indicator is not sufficiently stable, so we also introduce information density based on entropy calculation of neuronal activation as a discriminative factor. By calculating the information density of the first hidden layer for models with two-layer dimensions of $[512, 64]$ and $[256, 64]$ representing these two critical states, the average values are approximately 0.75 and 0.95, respectively. In Figure 3, we present the distribution of infinity norm of neuronal outputs for the selected layers of these two models, using kernel density estimation to demonstrate their probability density differences. It is evident that in models with more sufficient plasticity, neuronal outputs are more balanced (with peaks significantly lower than in the upper subplot). Models in this state typically possess stronger representational capacity, which validates the effectiveness of our established discriminative factors.\par
\begin{figure}[h]
	\centering
	\includegraphics[width=0.33\textwidth]{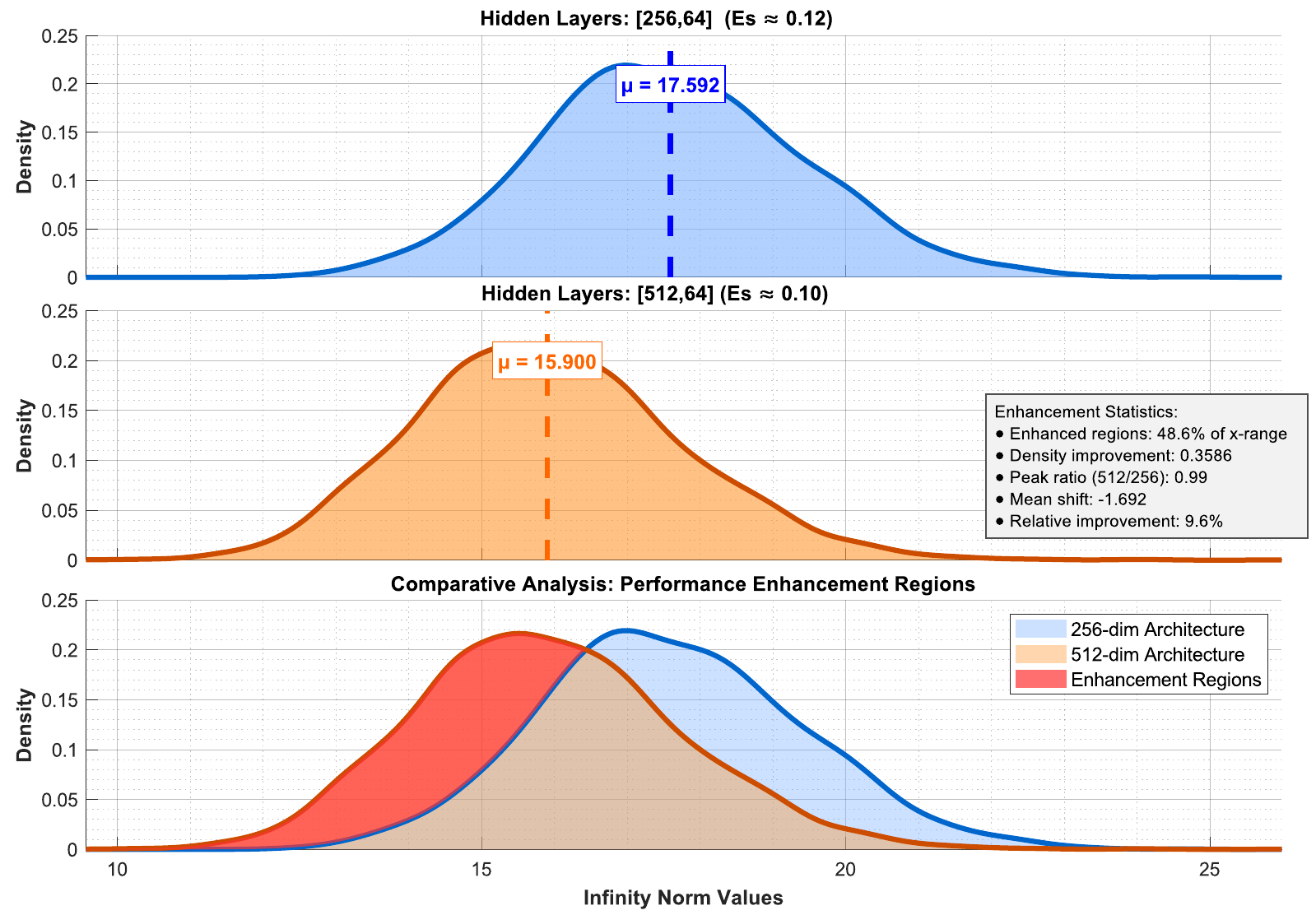} 
	\caption{Sufficient plasticity leads to more uniform output distribution which means greater information density growth potential.}
	\label{fig3}
\end{figure}

\begin{table*}[t]
	\centering
	\caption{Performance Comparison Across Different Incremental Learning Scenarios}
	\label{tab:incremental_comparison}
	\begin{tabular}{|l|l|c|c|c|c|}
		\hline
		\textbf{Scenario} & \textbf{Method} & \textbf{AA}$\uparrow$ & \textbf{BWT} $\uparrow$ closer to 0 & \textbf{FWT}$\uparrow$ & \textbf{FA}$\uparrow$ \\
		\hline
		\multirow{4}{*}{\begin{tabular}[c]{@{}l@{}}\textbf{Single Increment}\\ \textbf{(IPTAS-Tbps)}\\ \textbf{4+3 classes}\end{tabular}} 
		& Base & $0.814 \pm 0.052$ & $ -0.343 \pm 0.195$ & $0.804 \pm 0.077$ & $0.730 \pm 0.123$ \\
		& LwF & $0.819 \pm 0.070$ & $-0.429 \pm 0.311$ & $0.827 \pm 0.065$ & $0.722 \pm 0.156$ \\
		& EWC & $0.807 \pm 0.064$ & $-0.470 \pm 0.297$ & $\textbf{0.834} \pm \textbf{0.057}$ & $0.684 \pm 0.160$ \\
		& \cellcolor{lightgray!30}\textbf{PRIME (ours)} & \cellcolor{lightgray!30}$\textbf{0.856} \pm \textbf{0.117}$ & \cellcolor{lightgray!30}$\textbf{-0.140} \pm \textbf{0.391}$ & \cellcolor{lightgray!30}$0.689 \pm 0.137$ & \cellcolor{lightgray!30}$\textbf{0.787} \pm \textbf{0.246}$ \\
		\hline
		\multirow{4}{*}{\begin{tabular}[c]{@{}l@{}}\textbf{Low Task Similarity}\\ \textbf{(ISCX-VPN2016)}\end{tabular}} 
		& Base & $0.703 \pm 0.052$ & $-0.342 \pm 0.136$ & $\textbf{0.804} \pm \textbf{0.049}$ &$0.595 \pm 0.088$ \\
		& LwF &$0.716 \pm 0.125$ & $-0.361 \pm 0.076$ & $0.773 \pm 0.068$ & $0.585 \pm 0.178$ \\
		& EWC & $0.691 \pm 0.141$ & $-0.413 \pm 0.273$ & $0.775 \pm 0.051$ & $0.532 \pm 0.210$ \\
		& \cellcolor{lightgray!30}\textbf{PRIME (ours)} & \cellcolor{lightgray!30}$\textbf{0.775} \pm \textbf{0.098}$ & \cellcolor{lightgray!30}$\textbf{-0.165} \pm \textbf{0.167}$ & \cellcolor{lightgray!30}$0.700 \pm 0.100$ & \cellcolor{lightgray!30}$\textbf{0.703} \pm \textbf{0.159}$ \\
		\hline
		\multirow{4}{*}{\begin{tabular}[c]{@{}l@{}}\textbf{Continual Increment}\\ \textbf{(MIRAGE19)}\\ \textbf{10+2 per stage}\end{tabular}} 
		& Base & $0.642 \pm 0.032$ & $-0.412 \pm 0.043$ & $\textbf{0.837} \pm \textbf{0.020}$ & $0.572 \pm 0.063$ \\
		& LwF & $0.649 \pm 0.044$ & $-0.375 \pm 0.045$ & $0.835 \pm 0.017$ & $0.585 \pm 0.055$ \\
		& EWC & $0.662 \pm 0.038$ & $-0.349 \pm 0.031$ & $0.833 \pm 0.017$ & $0.605 \pm 0.065$ \\
		& \cellcolor{lightgray!30}\textbf{PRIME (ours)} & \cellcolor{lightgray!30}$\textbf{0.685} \pm \textbf{0.051}$ & \cellcolor{lightgray!30}$\textbf{-0.259} \pm \textbf{0.038}$ & \cellcolor{lightgray!30}$0.774 \pm 0.018$ & \cellcolor{lightgray!30}$\textbf{0.640} \pm \textbf{0.033}$ \\
		\hline
		\multicolumn{6}{@{}r@{}}{\vspace{2mm}\footnotesize{*All data presented as mean ± half-range}}
	\end{tabular}
\end{table*}

\begin{figure*}[t]
	\vspace{-5mm}
	\centering
	\includegraphics[width=0.9\textwidth]{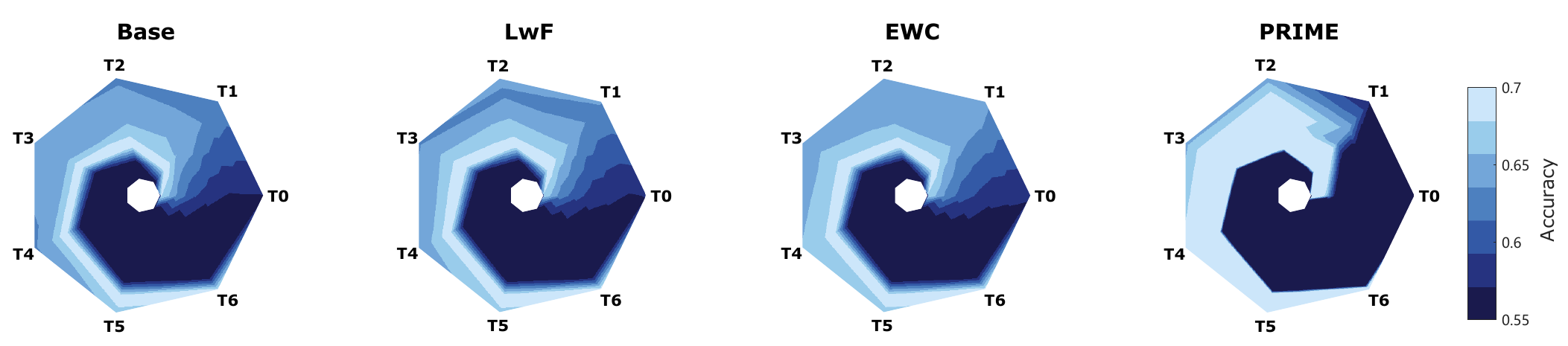} 
	\caption{Polar Visualization of Algorithm Performance of Continual Incremental Scenarios}
	\label{fig4}
\end{figure*}

In summary, combining the sensitivity to parameter expansion changes (the gap between information density trigger threshold and safety threshold is approximately 4 times that of the effective rank ratio) and the safety and trigger node values of the two factors, we can establish the plasticity determination formula: $\kappa_s=0.8pr_1+0.2pr_2$.\par

\subsection{Effect analysis under different incremental scenarios}
To evaluate the performance of continual learning algorithms, we adopt five widely-used metrics. Average Accuracy (AA) measures the overall performance across all tasks: 
$\frac{1}{T} \sum_{i=1}^{T} R_{i,i}$, where $R_{i,i}$ is the accuracy on $i_{th}$ task. Backward Transfer (BWT) quantifies the effect of learning new tasks on previous ones: $\frac{1}{T-1} \sum_{i=1}^{T-1} [R_{T,i} - R_{i,i}]$, where negative values indicate catastrophic forgetting. Forward Transfer (FWT) measures how prior learning benefits new tasks: $\frac{1}{T-1} \sum_{i=2}^{T} [R_{i-1,i} - R_{0,i}]$, where $R_{0,i}$ represents random initialization performance. Final Accuracy (FA) evaluates the model's final performance: $\frac{1}{T} \sum_{i=1}^{T} R_{T,i}$. Higher values are better for AA, FWT, and FA. For BWT, values close to zero or positive are preferred, as negative values indicate catastrophic forgetting.\par
We evaluate PRIME against established algorithms (LwF and EWC) on three datasets representing diverse incremental learning scenarios. IPTAS-Tbps uses a 4+3 class setup for single increment evaluation. ISCX-VPN2016 tests robustness under low task similarity conditions with minimal overlap between tasks. MIRAGE19 examines continual learning with 10 base classes plus 2 classes per subsequent stage.\par
Each dataset is randomly shuffled before train/test splitting, with multiple experimental repetitions to ensure robust comparisons across algorithms and scenarios ranging from simple task extensions to complex multi-stage continual learning.

\subsubsection{Single incremental scenarios with IPTAS-Tpds}
As shown in Table II, in the single increment learning scenario with IPTAS-Tbps dataset (4+3 classes), our proposed PRIME method demonstrates superior performance compared to baseline approaches. PRIME achieves the highest AA and FA, significantly outperforming Base, LwF, and EWC methods. Most notably, PRIME exhibits exceptional resistance to catastrophic forgetting with substantially better BWT performance compared to other methods. While EWC shows slightly better FWT capability, PRIME strategically trades off a portion of forward recognition accuracy for significant mitigation of catastrophic forgetting, resulting in better overall accuracy performance. This design choice allows PRIME to maintain competitive forward learning ability while achieving the best overall balance between learning new tasks and retaining previous knowledge.

\subsubsection{Low task similarity incremental scenarios with ISCX-VPN2016}
As shown in Table II, under the low task similarity scenario on ISCX-VPN2016, where we partition the application classification into four task types: "Chat", "Streaming", "File Transfer", and "Others", our proposed PRIME method significantly outperforms all baseline approaches. While traditional methods like EWC and LwF struggle in scenarios with low task similarity, exhibiting severe catastrophic forgetting, PRIME demonstrates superior knowledge retention with substantially improved backward transfer performance. Additionally, PRIME achieves the highest average accuracy and final accuracy, indicating its effectiveness in both learning new tasks and preserving previously acquired knowledge when tasks are dissimilar.

\subsubsection{Continual incremental scenarios with MIRAGE19}

As shown in Table II, in the continual increment scenario on MIRAGE19 with 10+2 classes per stage, PRIME consistently outperforms all baseline methods across key continual learning metrics. Our approach achieves the highest AA(0.685) and FA(0.640), demonstrating its effectiveness in incrementally learning new classes. Most importantly, PRIME exhibits significantly reduced catastrophic forgetting with a backward transfer of -0.259, substantially outperforming Base (-0.412), LwF (-0.375), and EWC (-0.349). While the forward transfer is slightly lower than baseline methods, PRIME's superior ability to retain knowledge from previous stages makes it particularly well-suited for practical continual increment scenarios where preserving learned representations is crucial. Besides, Figure 4 illustrates the continual learning performance comparison across four approaches. The baseline method (leftmost) shows significant performance degradation in later stages, evidenced by darker colors in the outer regions. The two comparative algorithms, LwF and EWC (second and third from left), demonstrate moderate improvement but still exhibit substantial forgetting. In contrast, our proposed PRIME algorithm (rightmost) maintains superior performance retention, characterized by predominantly lighter colors in the outer regions, indicating better resistance to catastrophic forgetting across sequential tasks.

\section{Related Work}

\subsection{Encrypted Traffic Classification}
Research on encrypted traffic classification has evolved from traditional techniques to more advanced methods due to the reduced availability of plain-text information caused by encryption protocols like TLS/SSL. Initially, port-based~\cite{2004Transport} and payload-based~\cite{2010Automatic} approaches utilized plain-text information for traffic classification, but the trend toward traffic encryption and the use of dynamic port mapping have rendered these methods ineffective. The focus has since shifted to machine learning and deep learning techniques. While early machine learning methods, such as decision trees and K-nearest neighbors~\cite{2009A}, offered some advancements, they largely relied on statistical features and struggled to effectively handle the sequential characteristics of network traffic.\par

The rise of deep learning has significantly enhanced traffic classification. Advanced architectures, such as residual networks~\cite{He_2016_CVPR}, facilitate deeper feature extraction. For example, Wei et al. transformed session-grouped packet data into image pixels for CNN classification~\cite{7899588}. Chen et al. developed the LS-LSTM model to analyze sequence features in encrypted traffic, achieving higher accuracy ~\cite{LS}. In recent developments, Dai applied the Transformer model from natural language processing to traffic classification in 2022~\cite{9916060}, and M Shen et al. successfully utilized graph neural networks for classifying encrypted traffic in 2021~\cite{9319399}.

\subsection{Incremental Learning}
Incremental learning aims for rapid adaptation to new tasks through model fine-tuning in the face of complex data changes. These methods typically leverage new and a small amount of old data to adjust certain parameters, which can lead to "catastrophic forgetting". This phenomenon occurs when a model improves on new data but performs poorly on old tasks due to excessive reliance on new data training. Commonly used strategies to mitigate this issue include regularization constraints, data replay, and parameter separation ~\cite{belouadah2020comprehensivestudyclassincremental}.\par
In a certain year, Zhizhong Li and Derek Hoiem proposed the LwF(Learning without Forgetting) algorithm ~\cite{8107520}, which constrains the deviation of old model predictions to avoid excessive updates. However, its effectiveness declines as task relevance weakens. Similarly, James Kirkpatrick et al. introduced the EWC method~\cite{EWC}, which incorporates parameter change penalties to prevent over-updating. Additionally, data replay methods, like iCaRL~\cite{Rebuffi_2017_CVPR}, select representative old data to avoid forgetting but require extra storage and computational resources. Variants using GANs~\cite{NIPS2017_0efbe980} or VAE~\cite{nguyen2017variational} substitute storage with computational costs without solving the underlying issue.\par
Another approach expands model size using gating neurons while keeping parameters for old tasks unchanged. While this mitigates forgetting, larger models consume more memory, affecting training and application. \par
Some traffic classification studies have utilized incremental learning methods, such as: MEMENTO~\cite{CERASUOLO2024110374}, CIL4EMTD~\cite{10858947}. However, these approaches only employ specific categories of the aforementioned incremental learning techniques.

\subsection{Plasticity in deep continual learning}
Shibhansh Dohare et al.~\cite{dohare2024loss} proposed research on model plasticity in incremental learning, discovering that certain model metrics are closely related to the learning effectiveness of models in incremental tasks. They summarized these metrics as model plasticity. Experiments showed that when model plasticity declines, there is a noticeable drop in performance across various tasks, including incremental updates in deep learning classification tasks, fitting tasks, and reinforcement learning-related tasks. 

\section{Conclusion and Future Work}
The proposed PRIME architecture innovatively introduces effective rank ratio and information density-based neuron inactivity assessment of the observation layer to determine model plasticity, i.e., expressive capacity. This approach is particularly suitable for encrypted traffic classification scenarios where samples evolve rapidly but computational resources are often constrained, enabling timely model expansion to achieve better efficiency.Across various incremental learning scenarios constructed from multiple datasets, PRIME achieves superior performance over mainstream algorithms across multiple metrics, requiring only 1-2 minimal-scale model expansions based on our observations.\par 
However, plasticity metrics introduce additional computational overhead. Our findings reveal complex correlations between plasticity variations and newly introduced task characteristics, requiring further investigation. Future research should focus on optimizing observation intervals, exploring alternative model expansion methods, and developing analytical frameworks for task-plasticity relationships.

\bibliographystyle{plain}
\bibliography{\jobname}

\begin{thebibliography}{10}

\bibitem{googletransparencyreport}
Google transparency report: Https overview.
\newblock \url{https://transparencyreport.google.com/https/overview}.
\newblock Accessed: 2023-10-01.

\bibitem{aceto2019mirage}
G.~{Aceto}, D.~{Ciuonzo}, A.~{Montieri}, V.~{Persico}, and A.~{Pescap{\`e}}.
\newblock Mirage: Mobile-app traffic capture and ground-truth creation.
\newblock In {\em IEEE 4th International Conference on Computing, Communication
  and Security (ICCCS 2019)}, Oct 2019.

\bibitem{distiller}
Giuseppe Aceto, Domenico Ciuonzo, Antonio Montieri, and Antonio Pescap{\'e}.
\newblock Distiller: Encrypted traffic classification via multimodal multitask
  deep learning.
\newblock {\em Journal of Network and Computer Applications}, 183:102985, 2021.

\bibitem{ISP}
Ali~Javed Azhar-ud din, Ayesha Hanif, M~Awais Azam, and Tasawer Hussain.
\newblock Development of postpaid and prepaid billing system for isps.
\newblock {\em Proceedings Appeared on IOARP Digital Library}, 2016.

\bibitem{malware}
Onur Barut, Yan Luo, Peilong Li, and Tong Zhang.
\newblock R1dit: Privacy-preserving malware traffic classification with
  attention-based neural networks.
\newblock {\em IEEE Transactions on Network and Service Management},
  20(2):2071--2085, 2022.

\bibitem{belouadah2020comprehensivestudyclassincremental}
Eden Belouadah, Adrian Popescu, and Ioannis Kanellos.
\newblock A comprehensive study of class incremental learning algorithms for
  visual tasks, 2020.

\bibitem{CERASUOLO2024110374}
Francesco Cerasuolo, Alfredo Nascita, Giampaolo Bovenzi, Giuseppe Aceto,
  Domenico Ciuonzo, Antonio Pescapè, and Dario Rossi.
\newblock Memento: A novel approach for class incremental learning of encrypted
  traffic.
\newblock {\em Computer Networks}, 245:110374, 2024.

\bibitem{chen2015net2net}
Tianqi Chen, Ian Goodfellow, and Jonathon Shlens.
\newblock Net2net: Accelerating learning via knowledge transfer.
\newblock {\em arXiv preprint arXiv:1511.05641}, 2015.

\bibitem{LS}
Zihan Chen, Guang Cheng, Zijun Wei, Ziheng Xu, Nan Fu, and Yuyang Zhou.
\newblock Higher layers, better results: Application layer feature engineering
  in encrypted traffic classification.
\newblock In Lei Wang, Michael Segal, Jenhui Chen, and Tie Qiu, editors, {\em
  Wireless Algorithms, Systems, and Applications}, pages 548--556, Cham, 2022.
  Springer Nature Switzerland.

\bibitem{chen2022a3c}
Zihan Chen, Guang Cheng, Ziheng Xu, Keya Xu, Yuhang Shan, and Jiakang Zhang.
\newblock A3c system: one-stop automated encrypted traffic labeled sample
  collection, construction and correlation in multi-systems.
\newblock {\em Applied Sciences}, 12(22):11731, 2022.

\bibitem{9916060}
Jianbang Dai, Xiaolong Xu, Honghao Gao, Xinheng Wang, and Fu~Xiao.
\newblock Shape: A simultaneous header and payload encoding model for encrypted
  traffic classification.
\newblock {\em IEEE Transactions on Network and Service Management},
  20(2):1993--2012, 2023.

\bibitem{dohare2024loss}
Shibhansh Dohare, J~Fernando Hernandez-Garcia, Qingfeng Lan, Parash Rahman,
  A~Rupam Mahmood, and Richard~S Sutton.
\newblock Loss of plasticity in deep continual learning.
\newblock {\em Nature}, 632(8026):768--774, 2024.

\bibitem{VPN2016}
Gerard Draper-Gil, Arash~Habibi Lashkari, Mohammad Saiful~Islam Mamun, and
  Ali~A Ghorbani.
\newblock Characterization of encrypted and vpn traffic using time-related.
\newblock In {\em Proceedings of the 2nd international conference on
  information systems security and privacy (ICISSP)}, pages 407--414, 2016.

\bibitem{vpn}
Lulu Guo, Qianqiong Wu, Shengli Liu, Ming Duan, Huijie Li, and Jianwen Sun.
\newblock Deep learning-based real-time vpn encrypted traffic identification
  methods.
\newblock {\em Journal of Real-Time Image Processing}, 17(1):103--114, 2020.

\bibitem{Gbps}
Simon Hauger, Thomas Wild, Arthur Mutter, Andreas Kirstaedter, Kimon Karras,
  Rainer Ohlendorf, Frank Feller, and Joachim Scharf.
\newblock Packet processing at 100 gbps and beyond - challenges and
  perspectives.
\newblock In {\em 2009 ITG Symposium on Photonic Networks}, pages 1--10, 2009.

\bibitem{He_2016_CVPR}
Kaiming He, Xiangyu Zhang, Shaoqing Ren, and Jian Sun.
\newblock Deep residual learning for image recognition.
\newblock In {\em Proceedings of the IEEE Conference on Computer Vision and
  Pattern Recognition (CVPR)}, June 2016.

\bibitem{2009A}
Shijun Huang, Kai Chen, Chao Liu, Alei Liang, and Haibing Guan.
\newblock A statistical-feature-based approach to internet traffic
  classification using machine learning.
\newblock In {\em Proceedings of the International Conference on Ultra Modern
  Telecommunications, ICUMT 2009, 12-14 October 2009, St. Petersburg, Russia},
  2009.

\bibitem{2004Transport}
Thomas Karagiannis, Andre Broido, Michalis Faloutsos, and Kc~Claffy.
\newblock Transport layer identification of p2p traffic.
\newblock In {\em ACM SIGCOMM conference on Internet measurement}, 2004.

\bibitem{forgetting}
Ronald Kemker, Marc McClure, Angelina Abitino, Tyler Hayes, and Christopher
  Kanan.
\newblock Measuring catastrophic forgetting in neural networks.
\newblock {\em Proceedings of the AAAI Conference on Artificial Intelligence},
  32(1), Apr. 2018.

\bibitem{khanna2006application}
Gunjan Khanna, Kirk Beaty, Gautam Kar, and Andrzej Kochut.
\newblock Application performance management in virtualized server
  environments.
\newblock In {\em 2006 IEEE/IFIP Network Operations and Management Symposium
  NOMS 2006}, pages 373--381. IEEE, 2006.

\bibitem{EWC}
James Kirkpatrick, Razvan Pascanu, Neil Rabinowitz, Joel Veness, Guillaume
  Desjardins, Andrei~A. Rusu, Kieran Milan, John Quan, Tiago Ramalho, Agnieszka
  Grabska-Barwinska, Demis Hassabis, Claudia Clopath, Dharshan Kumaran, and
  Raia Hadsell.
\newblock Overcoming catastrophic forgetting in neural networks.
\newblock {\em Proceedings of the National Academy of Sciences},
  114(13):3521--3526, 2017.

\bibitem{10858947}
Jiarui Li, Lei Du, Yilu Chen, Liyi Zeng, Hao Li, and Zhaoquan Gu.
\newblock Cil4emtd: A novel class incremental learning method for encrypted
  malware traffic detection.
\newblock In {\em 2024 IEEE 9th International Conference on Data Science in
  Cyberspace (DSC)}, pages 300--307, 2024.

\bibitem{li2006matrix}
Ren-Cang Li.
\newblock Matrix perturbation theory.
\newblock {\em Handbook of linear algebra}, pages 15--21, 2006.

\bibitem{8107520}
Zhizhong Li and Derek Hoiem.
\newblock Learning without forgetting.
\newblock {\em IEEE Transactions on Pattern Analysis and Machine Intelligence},
  40(12):2935--2947, 2018.

\bibitem{lin2022bert}
Xinjie Lin, Gang Xiong, Gaopeng Gou, Zhen Li, Junzheng Shi, and Jing Yu.
\newblock Et-bert: A contextualized datagram representation with pre-training
  transformers for encrypted traffic classification.
\newblock In {\em Proceedings of the ACM Web Conference 2022}, pages 633--642,
  2022.

\bibitem{drift}
Navid Malekghaini, Elham Akbari, Mohammad~A. Salahuddin, Noura Limam, Raouf
  Boutaba, Bertrand Mathieu, Stephanie Moteau, and Stephane Tuffin.
\newblock Deep learning for encrypted traffic classification in the face of
  data drift: An empirical study.
\newblock {\em Computer Networks}, 225:109648, 2023.

\bibitem{nguyen2017variational}
Cuong~V Nguyen, Yingzhen Li, Thang~D Bui, and Richard~E Turner.
\newblock Variational continual learning.
\newblock {\em arXiv preprint arXiv:1710.10628}, 2017.

\bibitem{Rebuffi_2017_CVPR}
Sylvestre-Alvise Rebuffi, Alexander Kolesnikov, Georg Sperl, and Christoph~H.
  Lampert.
\newblock icarl: Incremental classifier and representation learning.
\newblock In {\em Proceedings of the IEEE Conference on Computer Vision and
  Pattern Recognition (CVPR)}, July 2017.

\bibitem{7098875}
Olivier Roy and Martin Vetterli.
\newblock The effective rank: A measure of effective dimensionality.
\newblock In {\em 2007 15th European Signal Processing Conference}, pages
  606--610, 2007.

\bibitem{LLMnetworking}
Jiawei Shao and Xuelong Li.
\newblock Ai flow at the network edge.
\newblock {\em IEEE Network}, pages 1--1, 2025.

\bibitem{9319399}
Meng Shen, Jinpeng Zhang, Liehuang Zhu, Ke~Xu, and Xiaojiang Du.
\newblock Accurate decentralized application identification via encrypted
  traffic analysis using graph neural networks.
\newblock {\em IEEE Transactions on Information Forensics and Security},
  16:2367--2380, 2021.

\bibitem{NIPS2017_0efbe980}
Hanul Shin, Jung~Kwon Lee, Jaehong Kim, and Jiwon Kim.
\newblock Continual learning with deep generative replay.
\newblock In I.~Guyon, U.~Von Luxburg, S.~Bengio, H.~Wallach, R.~Fergus,
  S.~Vishwanathan, and R.~Garnett, editors, {\em Advances in Neural Information
  Processing Systems}, volume~30. Curran Associates, Inc., 2017.

\bibitem{NIPS2017_3f5ee243}
Ashish Vaswani, Noam Shazeer, Niki Parmar, Jakob Uszkoreit, Llion Jones,
  Aidan~N Gomez, \L~ukasz Kaiser, and Illia Polosukhin.
\newblock Attention is all you need.
\newblock In I.~Guyon, U.~Von Luxburg, S.~Bengio, H.~Wallach, R.~Fergus,
  S.~Vishwanathan, and R.~Garnett, editors, {\em Advances in Neural Information
  Processing Systems}, volume~30. Curran Associates, Inc., 2017.

\bibitem{7899588}
Wei Wang, Ming Zhu, Xuewen Zeng, Xiaozhou Ye, and Yiqiang Sheng.
\newblock Malware traffic classification using convolutional neural network for
  representation learning.
\newblock In {\em 2017 International Conference on Information Networking
  (ICOIN)}, pages 712--717, 2017.

\bibitem{2010Automatic}
Yu~Wang, Yang Xiang, and Shun~Zheng Yu.
\newblock Automatic application signature construction from unknown traffic.
\newblock In {\em IEEE International Conference on Advanced Information
  Networking \& Applications}, 2010.

\bibitem{Rosetta}
Renjie Xie, Yixiao Wang, Jiahao Cao, Enhuan Dong, Mingwei Xu, Kun Sun, Qi~Li,
  Licheng Shen, and Menghao Zhang.
\newblock Rosetta: Enabling robust tls encrypted traffic classification in
  diverse network environments with tcp-aware traffic augmentation.
\newblock In {\em Proceedings of the ACM Turing Award Celebration Conference -
  China 2023}, ACM TURC '23, page 131–132, New York, NY, USA, 2023.
  Association for Computing Machinery.

\end{thebibliography}

\end{document}